\documentclass[12pt,preprint]{aastex}
\usepackage{natbib}
\shorttitle{Investigation of the New Local Group Galaxy VV 124}
\shortauthors{{N. A. Tikhonov, S. N. Fabrika, O. N. Sholukhova, and A. I. Kopylov}}
\begin{document}
\makeatletter


\fontsize{12}{12} \selectfont

\title{Investigation of the New Local Group Galaxy VV 124}
\author{{N. A. Tikhonov, S. N. Fabrika, O. N. Sholukhova, and A. I. Kopylov}}
\affil{Special Astrophysical Observatory, Russian Academy of Sciences, N.Arkhyz, KChR, 369167, Russia}

\begin{abstract}
We present the results of our stellar photometry and spectroscopy for the new Local Group
galaxy VV~124 (UGC 4879) obtained with the 6-m BTA telescope. The presence of a few bright supergiants
in the galaxy indicates that the current star formation process is weak. The apparent distribution of
stars with different ages in VV 124 does not differ from the analogous distributions of stars in irregular
galaxies, but the ratio of the numbers of young and old stars indicates that VV~124 belongs to the rare
Irr/Sph type of galaxies. The old stars (red giants) form the most extended structure, a thick disk with an
exponential decrease in the star number density to the edge. Definitely, the young population unresolvable
in images makes a great contribution to the background emission from the central galactic regions. The
presence of young stars is also confirmed by the [O III] emission line visible in the spectra that belongs to
extensive diffuse galactic regions. The mean radial velocity of several components (two bright supergiants,
the unresolvable stellar population, and the diffuse gas) is $v_h = -70 \pm 15$\,km s$^{-1}$
and the velocity with which VV~124 falls into the Local Group is  $v_{LG} = -12 \pm 15$\,km s$^{-1}$. 
We confirm the distance to the galaxy ($D=1.1\pm 0.1$~Mpc) and the metallicity of red giants ([Fe/H] = $-$1.37) 
found by Kopylov et al. (2008).
VV 124 is located on the periphery of the Local Group approximately at the same distance from M~31 and
our Galaxy and is isolated from other galaxies. The galaxy LeoA nearest to it is 0.5 Mpc away.

\end{abstract}

\section*{INTRODUCTION}
Our Local Group holds a special place in studying the parameters of nearby groups of galaxies, because
all of its galaxies are located at the shortest distances from our Galaxy. Such a close neighborhood gives
a unique opportunity to study in detail the stellar population of individual galaxies and the kinematics
of their subsystems as well as to accurately measure the distances to individual galaxies and their velocities.
 The values obtained are used to determine the dynamical characteristics of the Local Group and to
search for possible correlations between the parameters of an individual galaxy and its location among the
Local Group galaxies. When the groups of galaxies are studied, the Local Group parameters are commonly
 used as a reference in comparing them with the parameters of morphologically similar groups of galaxies.

It is not always obvious whether a galaxy belongs to the Local Group. This is particularly true of pe-
ripheral galaxies. For example, the list of Local Group galaxies from Mateo (1998) numbers 40 members,
while the more recent paper by Van den Bergh (2000) contains only 36 members, because some of the dis-
tant galaxies were excluded from the list: UGCA~92, UGCA~438, GR~8, and others. Refined data show that
these galaxies are actually outside the Local Group. In the long run, the size and mass of the Local Group
and, hence, the membership of peripheral galaxies in the Local Group are determined from accurate data
on the velocities and distances of all the galaxies that constitute the Local Group \citep{whi05}.

In searching for new nearby galaxies, great hopes have been pinned on the Sloan Digital Sky Survey
(SDSS). Indeed, when approximately a quarter of the celestial sphere was investigated, more than ten new
Local Group galaxies were found \citep{zuc04, zuc06, zuc07, wil05, wal07, bel06, irw07}.
 Three new Local Group galaxies have recently been discovered
in $CFHT$ images when the southwestern quadrant of M31 was surveyed \citep{mcc08}.

All of the galaxies discovered in the last decade have low masses and low luminosities ($M_B = -3\div  -9$).
 Therefore, the discovery that the well-known bright irregular galaxy VV~124 is a member of the
Local Group \citep{kop08} with luminosity $M_B = -12$ was quite unexpected. The interest in
the new Local Group member VV~124 is that this galaxy is fairly isolated from the neighboring galaxies
and it has evolved without any interaction with the giant M31 or our Galaxy. Since VV~124 is located
on the periphery of the Local Group, an accurate measurement of its main characteristics and, most
importantly, its radial velocity allows the location of the zero-velocity sphere, i.e., the dynamical boundary
of the Local Group, to be refined. The first publication \citep{kop08} described the history of the
discovery of VV~124 and briefly presented our preliminary results. In this paper, we present more complete
results of our analysis of the observations for VV~124. Since the radial velocity estimate for the new galaxy
is a fundamentally important characteristic because of its location at the boundary of the Local Group,
we present reprocessed (to increase the accuracy) and more detailed radial velocity measurements for
various components of VV~124.

\section*{OBSERVATIONS}
The observations of VV 124 and their reduction were described in detail previously \citep{kop08}; these were carried out with the 6-m BTA telescope on January 11, 2008, (direct $V$
and $I_c$ images) and on February 6, 2008 (long-slit spectroscopy). Since the galaxy is well resolved
into stars in all images (Fig. 1), we could perform stellar photometry and, based on the results obtained,
determine the distance to the galaxy and study the distribution of stars with different ages over the
galactic body.

The wavelength range from 4000 to 5600 \AA \ was observed at a spectral resolution of 5~\AA.\ The initial
accuracy of constructing the wavelength scale is 5--10\,km\,s$^{-1}$. We checked the radial velocities in
the entire spectral range using atmospheric lines ([O\,I]\,$\lambda 5577.35$ and 
Hg\,I\,$\lambda 4046.56, 4358.34, 5460.74$);
this check showed the wavelength scale to be accurate to within 5\,km\,s$^{-1}$. However, the accuracy of
radial velocity measurements is also determined by the spectrum quality. The radial velocities of narrow
emission lines from nebulae are measured more reliably than those of broad hydrogen lines from the
background of unresolved stars. Below, we describe in detail the results of our velocity measurements for
various components of the galaxy. In this paper, we present reprocessed spectroscopic measurements. In
addition to the data published in Kopylov et al. (2008), we measured the velocities of a large number of
galactic components distributed more widely over the galactic body and, hence, these new measurements
well represent the velocity of VV~124. 

We used two spectrograph slit positions (see Fig. 5 below): the slit passed through the stars bl1 (slit 1)
and bl2 (slit 2) at the first and second positions, respectively. The slit width in both cases was $1\arcsec$. A weak
extended emission in the [OIII]~$\lambda$ 4959 and 5007 \AA,\ lines and in the H$\beta$ line was recorded at the two slit
positions. A very compact H II region emitting the H$\beta$ line was detected on slit 1 and a distant emission-line
 galaxy was projected by chance onto slit 2.

\section*{STELLAR PHOTOMETRY}

The stellar photometry was performed in a standard way using DAOPHOT II in MIDAS \citep{ste94}.
 As a result, we obtained tables with stellar coordinates, magnitudes, and associated parameters
 that allowed us to estimate the photometric accuracy and the deviation of the profile of each
photometered object from the standard profile of a PSF star (CHI and SHARPNESS). The table
was selected by these parameters (CHI$ < 1.3$ and $-0.3 <$ SHARPNESS $< 0.3$) \citep{ste94} to remove
 all of the diffuse objects and false stars that appeared in the list due to a few cosmic-ray particle
hits remained in the images after the primary data reduction from the final list of stars.
The transformation equations from the instrumental magnitudes obtained by the photometry program
to the $V$ and $I$ magnitudes of the Kron--Cousins system were derived from the photometry of standard
stars in field SA98 \citep{lan92}, whose images were obtained on the same night as the images of the
galaxy VV 124 under study:

 $$ (V-I)_{c} = 0.954 \cdot (v_{0} - i_{0}) + 0.881 \eqno(1)$$
 $$ I_{c} = i_{0} + 0.051 \cdot (V-I) + 25.565  \eqno(2)$$

The results of our photometry are presented in Fig. 2 in the form of a Hertzsprung--Russell (or CM)
diagram. The constructed diagram does not differ from the analogous diagrams for galaxies with low
star formation rates (Irr/Sph), as is observed, for example, in the Phoenix galaxy \citep{hol00}.
We see from the diagram in Fig. 2 that most of the stars in the galaxy are old red giant branch (RGB)
stars, while the blue supergiants form a sparsely populated branch due to the weakness of the star 
formation process. The two fitted isochrones with metallicity Z = 0.008 and ages of 16 and 63 Myr (\citep{ber94}
 indicate that the two blue stars bl1 and bl2 whose spectra are described below are no older than
20 Myr. Our spectroscopy (see below) shows that bl1 is a blue supergiant and bl2 is a yellow supergiant.
Thus, weak star formation took place in VV 124 several tens of Myr ago.

Since the RGB is clearly seen on the СМ diagram (Fig. 2), we used the TRGB method \citep{lee93} 
based on determining the tip of the RGB to estimate the distance. The accuracy of the method
increases if only the galaxy's peripheral stars are used for the measurements, with the central regions ex-
cluded. This is because the spatial distributions of stars with different ages in irregular galaxies differ 
\citep{tik05, tik06}. By their presence, the brighter (than red giants) AGB stars on the CM diagram smear
the sharp tip of the RGB in the luminosity function. Since the AGB stars are strongly concentrated to the
 galactic center, their influence can be reduced by excluding the central galactic region from the measurements.
 Thus, in constructing the luminosity function of red giants and determining the tip of the RGB (TRGB jump),
 we used a spatial selection of stars along the radius of the galaxy and along its minor axis, leaving the
 stars at  $Rad > 90\arcsec$ and  $-100\arcsec < b < 100\arcsec$.

The restrictions along the minor axis $b$ stem from the fact that the galaxy has well-defined boundaries and
 we reduce the number of background stars in the luminosity function, establishing this selection.
The luminosity function for red giants exhibits a sharp increase in the number of stars at
 $I = 21.2 \pm 0.1$ (Fig. 3), which corresponds to the beginning of the RGB (TRGB jump). Using the
inferred tip of the RGB, we determined the mean color indices of the giant branch: $(V-I)_{-3.5} = 1.45$ and
$(V-I)_{TRGB} = 1.65$ \citep{lee93}. The extinction toward VV 124 was taken from the measurements
by Schlegel et al. 1998): $A_V = 0.050$, $A_I$ = 0.029, and $E(V-I)=0.02$. At the adopted extinction coeffcients,
 we obtain: $I^0_{TRGB} = 21.17$, $(V-I)^0_{-3.5} = 1.43$, аnd $(V-I)^0_{TRGB} = 1.63$. Using equations from Lee
et al. (1993), we determined the metallicity of red giants for VV~124, [Fe/H] = $-$1.37, and the distance 
to the galaxy, $D = 1.1\pm 0.1$ Mpc. Since there is virtually no stellar metallicity gradient in dwarf
galaxies, [Fe/H] = $-$1.37 may be considered to be the metallicity of red giants in the entire galaxy.
 Thus, we confirm the distance to VV 124 and the metallicity of red giants obtained by Kopylov et al. (2008).

\section*{THE APPARENT DISTRIBUTION OF STARS WITH DIFFERENT AGES}

Having identified the stars with different ages on the CM diagram (Fig. 2): young stars (blue supergiants),
 intermediate-age stars (AGB), and old stars (red giants), we constructed their apparent distribution in 
the galaxy. Figure 4 shows the number density distribution of stars with different ages along
the minor axis. Along the major axis, the galaxy goes beyond our images. However, an extrapolation
is possible to determine its size, because the change in the number density of stars is usually expressed by
a linear dependence in logarithmic coordinates. 

Our measurements showed that the young blue stars in VV~124 form a  $0.76\times 0.47$ kpc subsystem
that is usually called a thin disk, the intermediate-age (AGB) stars form a $1.13\times 0.85$~kpc disk, and the
red giants form a $2.1 \times 1.3$~kpc ($6.\arcmin5 \times 4.\arcmin2$) stellar subsystem
 called a thick disk. We see from Fig.~4 that
the number density decreases exponentially toward the edge for all types of stars. This is obvious for the
red giants, while for the AGB and blue stars their distributions are consistent with exponential ones.
The only region of weak star formation is located in the galaxy asymmetrically relative to the apparent
optical galactic center coincident with the center of the distribution of red giants \citep{kop08}.
The ratio of the sizes of the thick (the region of red giants) and thin (the region of young stars) disks
is 2.8 and corresponds to the mean value for irregular galaxies \citep{tik05, tik06}. However, since the
number of young supergiants is small, VV~124 can be attributed to intermediate-type galaxies, between
irregular and spheroidal ones (Irr/Sph).

\section*{SPECTROSCOPY}
We measured the radial velocities of the two bright supergiants bl1 and bl2, the diffuse [O III] $\lambda$ 5007 \AA \
line emission, the total emission from unresolvable stars in VV 124, and the compact weak H II region.
The positions of all the objects studied are shown in Fig. 5. The nebular diffuse [O III] emission was
measured in regions {\it а} and {\it b} on slit 2 and in region {\it а} on slit 1. For a distance to the galaxy of 1.1 Mpc and
a scale of 5.3 pc/$\arcsec$, the sizes of the galactic regions where the radial velocities were measured
are about 330 pc in projection onto the galaxy. The projected separation between bl1 and bl2 is 110 pc.

The spectra of bl1 and bl2 are presented in Fig. 6. The narrow lines in the spectra indicate that these
stars are actually supergiants in VV 124. We estimated the temperatures of these stars by comparing their
 spectra with stellar atmosphere models from Munari et al. (2005)(http://archives.pd.astro.it/2500-10500/).
 The temperature and metallicity of bl2 were found to be $6000 \pm 100$\,K and [M/H]$=-0.5 \pm0.5$, respectively.
 The temperature of bl1 is $20000 \pm 3000$\,K. Only the hydrogen and He\,I lines are clearly seen in its spectrum,
 but weak metal lines are also noticeable. It can be concluded from the metal line intensities that the 
metallicity of this star is [M/H]$= 0.0 \div -0.5$, i.e., it is nearly solar.

We found the magnitudes and colors for bl1 and bl2: $V = 18.53$ and $(V-I) = - 0.13$ for bl1,
$V = 17.81$ and $(V-I) = +0.30$ for bl2. Using the previously determined distance modulus for VV~124,
we calculated the absolute magnitudes of bl1 and bl2: $M_v = -6.73$ and $M_v = -7.45$, respectively. The luminosities
 of these stars indicate that they are actually bright supergiants.

The hydrogen (H$\delta$, H$\gamma$ and H$\beta$) lines were measured in the spectra of bl1 and bl2, since these are the
deepest lines that give the highest measurement accuracy. The radial velocities were measured by fitting
a Gaussian into the hydrogen line cores. The radial velocities of bl1 and bl2 are $-90 \pm 15$\,km s$^{-1}$ and
 $-82 \pm 15$\,km s$^{-1}$, respectively. A fainter red stars is located near bl2 \citep{kop08}. 
The possibility that bl1 and bl2 have close neighbors cannot be ruled out, given the scale of 5.3 pc/$\arcsec$.
 However, the close velocities of both stars indicate that the results of our measurements are correct. Note
 that \citep{kop08}, we provided slightly different velocities of bl1 and bl2, differing by 11 and 7 km s$^{-1}$,
 respectively. These differences are within the limits of the measurement errors. We obtained the new velocities
of the stars after reprocessing the spectra and refining the measurements by comparing the observed line
profiles with those of model atmospheres for given temperatures.

The spectra of the galactic background produced by unresolvable stars were measured in regions {\it a + b}
on slit 1 and in region {\it b} on slit 2 (Fig. 5). The spectrum of the compact H II region was masked when
the spectra of these regions were measured, because the emission from this region distorted the hydrogen
absorption lines. Since the most intense emission of the unresolvable stellar population in VV 124 comes
from these regions (Fig. 5), we obtained fairly accurate radial velocity measurements for them. The
spectra of unresolvable stars contain three deep hydrogen lines, yet the accuracy of measuring the radial
velocity from these spectra is lower than that from the spectra of bl1 and bl2 due to the lower signal-to-
noise ratio. We determined the mean radial velocity of the stars in these regions from three hydrogen lines,
$-70 \pm 15$\,km s$^{-1}$. The stellar H$\beta$ absorption line was distorted by the weak emission line from the diffuse
emission of H II regions, which introduced an error into the radial velocity measurements. The diffuse H$\beta$
emission line observed in the spectra is redshifted (see below) relative to the absorption line of background
stars. Note that not only in the central galactic region but also along the entire slit (the slit length is 5 $\arcmin$ or
1.5 kpc), the main lines in the spectrum are hydrogen ones. This means that the young stellar population,
at least in the form of faint and unresolvable (in BTA images) stars, occupies larger volumes of the galaxy
than the star-forming region or even the thin disk, whose sizes were estimated from the distribution of
brighter stars.

There is a very weak diffuse emission in the [OIII]$\lambda$4959 and 5007 \AA\  and H$\beta$ emission lines in regions
 {\it a}
 and {\it b} on slit 2 and in region {\it a} on slit 1 (Fig. 5). Only the intense [O III] $\lambda$5007 \AA \ line
 is measurable,
while H$\beta$ is located in the broad absorption line of the stellar background. Figure 6 presents the spectrum
taken in a 13$\arcsec$-long segment of region {\it b} on slit 2 located from the western edge of region {\it b} (in such a
way that the segment did not include bright stars) to the group of stars containing the compact H II region.
Since this spectrum clearly shows a narrow [O III] line, accurate radial velocity measurements can be
performed. The measured radial velocity of the [O III] line on slit 2 is $-71 \pm 10$\,km s$^{-1}$ in region {\it a} and
$-47 \pm 15$\,km s$^{-1}$ in the segment of region {\it b}. The fluxes in the [O III]  $\lambda$5007 \AA\ line from 
one square
arcsec in these regions are $(7.1 \pm 1.0) \cdot 10^{-17}$ and $(7.4 \pm 1.1)\cdot 10^{-17}$ \,erg cm$^2$ s$^{-1}$. 
The radial velocity of the  $\lambda$5007 \AA\ line on slit 1 in region {\it a} is  $-54 \pm 15$\,km s$^{-1}$ and the flux in this line from one
$-54 \pm 15$\,km s$^{-1}$ and the flux in this line from one  square arcsec is  $(6.7 \pm 1.4) \cdot 10^{-17}$\, erg cm$^{-2}$s$^{-1}$.

The estimated radial velocities strongly suggest that the diffuse H II region emitting an intense
[O III] line belongs to VV~124. This region may be even more extended than the region of our 
measurements based on two slit positions. On slit 2, the [O III] emission begins slightly
 westward of bl2 and ends at several arcseconds from the compact H II region
(Fig.~5). The total size of the diffuse [O III] region along slit 2 is 32$\arcsec$ (170 pc). On slit 1, the [O III]
emission is observed between bl1 and the compact H II region. The size of the [O III] region along slit 1
is 14$\arcsec$ (75 pc). In the images of our long-slit spectra and on the diagrams of the [O III] line flux distribution
along the slits, we clearly see that the [O III] emission is nonuniform both spatially and in radial velocities.

Our spectroscopic measurements of the faint and compact H II region (Fig. 5) emitting in the H$\beta$  line
showed its size to be about 2$\arcsec$--2.5$\arcsec$ (11--13 pc) along slit 1 and no more than 1.7$\arcsec$~(9 pc)
 along slit 2. A certain structure of this H II region is noticeable in the image of the long-slit spectrum 
on slit 1. It is highly
likely that this compact H II region has a complex structure, both spatially and in radial velocities, but
the quality of the spectra obtained is too low for it to be studied in more detail. Since this region is located
amidst the bright background of stars and the H$\beta$ emission line lies in the deep absorption line of the
stellar background, the measurements of its radial velocity are not very reliable. The radial velocity of
this compact H II region is $-$36\,km s$^{-1}$ on slit 1 and $-$55\,km s$^{-1}$ on slit 2. There may be a systematic error
in these measurements, because the profile of the H${\beta}$ absorption line from the stellar background is asymmetric.
 Recall that the radial velocity of unresolvable stars in this galactic region is $-$70\,km s$^{-1}$.

Recently, VV\,124 has been observed by Oosterloo (2010) in the 21-cm line (Westerbork Synthesis
Radio Telescope). The galaxy turned out to be richin neutral hydrogen, with the region of its maximum
emission being located near bl2. This region of maximum emission of the hydrogen cloud is elongated
along the galaxy's major axis. The entire H I cloud is fairly extended, its sizes (at the photometric limit
of radio observations) are at least $2\arcmin\times 3\arcmin$, and it goes far outside the galaxy's optical body in the
 southeastern direction. The direction of the cloud elongation roughly coincides with the direction of the apparent 
distribution of bright young stars (almost along slit 2). The H I radial velocity in the region of maximum
 emission (near bl2) is $-$25 km s$^{-1}$ with a dispersion of several km s$^{-1}$, but in the southeastern part of
the cloud the H I radial velocity reaches $-$50 km s$^{-1}$. The total velocity range of the entire H\,I cloud is fairly
wide, from $-$50 to 0 km s$^{-1}$. It is highly likely that we observe the interaction of the H\,I cloud with VV 124
\citep{oos10} or, to be more precise, the neutral hydrogen cloud falls into the galaxy.

The fact that the extended and distant (from the galactic center) regions of the hydrogen cloud have a
radial velocity of $-$50 km s$^{-1}$ and the high total H I velocity dispersion confirm the results of our velocity
measurements for the galactic body. The estimated parameters of the spatial stellar and gaseous structure of 
VV\,124 make this galaxy an interesting object for more detailed studies.

Obviously, the H I cloud is located in the galaxy's field of attraction. Either it was captured by VV 124
and, in this case, has a ,,primordial'' origin or it is composed of the gas that was ejected from the galaxy
in the past during a powerful starburst. The radial velocity of the galaxy determined from unresolved
stars is $-$70 km s$^{-1}$, while the brightest part of the H\,I cloud located near bl2 has a radial velocity of
$-$25 km s$^{-1}$. The H I cloud most likely interacts with the galaxy and the place of contact between the cloud
and the galaxy is near bl2. Slow but extensive star formation that we observe in the form of young stars
and vast H\,II regions illuminated by them arises at the place of contact. Note that the [O III] gas velocity
dispersion is also high, from $-$47 to $-$71 km s$^{-1}$, which appreciably exceeds the measurement errors.

The characteristic accretion time of the H\,I cloud onto the galaxy estimated as the mean size ($2\arcmin$ in
projection) divided by the difference in the velocities of the distant parts of the cloud ($-$50 km s$^{-1}$ ) and
the galactic body ($-$70 km s$^{-1}$ ) is $3 \cdot 10^7 \div 10^8$\,yr. The mass of the central part of the galaxy 150 pc
in radius (the distance from the center to the H I emission maximum) was roughly estimated from the
difference of the H\,I velocities at the place of contact ($-$25 km s$^{-1}$ ) and the galactic body ($-$70 km s$^{-1}$) to
be $\sim 7 \cdot 10^7 M_{\sun}$.

A distant emission-line galaxy fell by chance on slit 2 (Fig. 5). Its coordinates (2000.0) are:
$\alpha = 09^h16^m02^s.5, \delta = +52^{o}50$\arcmin$ 23.5$\arcsec$ $.
 The galaxy's spectrum exhibits an intense and narrow (FWHM $< 6$ \AA)\ emission line at a
 wavelength $\lambda \approx 5124$ (Fig. 6). This is most likely the [O II] $ \lambda $3727 line at redshift
z = 0.375. At such z, the observed broad absorption line at a wavelength of $\sim 5640$ \AA\ can be interpreted
as H$\delta$, but this absorption line lies at the edge of our spectral range ($4000 - 5700$ \AA).\ In addition to the
intense and narrow emission line, narrow but several times weaker emission lines, for example, at $\sim 5060$ \AA,\
are also observed, but their intensities are comparable to noise.

\section*{VV 124(UGC 4879) AND THE LOCAL GROUP}

Figures 7 and 8 present the distributions of galaxies from the Local Group and its immediate neighborhoods
 in projection onto the Supergalactic plane and perpendicular to this plane. Whereas the galaxies VV~124
 and LeoA in Fig.~7 seem to be located near the small NGC~3109 group, we see from Fig.~8
that VV~124 and Leo~A lie in the flattened disk of Local Group galaxies, while the NGC~3109 group
lies at a considerable distance from the disk. The velocity of VV~124 relative to the center of mass of
the Local Group, which we take to be at an equal distance between М~31 and our Galaxy, is $v_{LG} = -12\pm 15$\,km s$^{-1}$.
Clearly, at such a velocity, VV~124 has never approached massive Local Group galaxies and
it may have entirely evolved without any strong gravitational interaction with massive galaxies. Such
 isolation of VV~124 makes the study of its star formation history particularly interesting, but
this requires observations with space telescopes.

\section*{DISCUSSION AND CONCLUSIONS}
In the history of the discovery of new Local Group galaxies, the times when bright Irr and Irr/Sph 
galaxies comparable in brightness to VV 24 were discovered have long been passed. Two such galaxies
(DDO210 and PegDIG) were discovered 50 years ago \citep{van59} and one galaxy (SagDIG)
was discovered 30 years ago \citep{ces77} before the appearance of the largest telescopes and
digital sky surveys. To all appearances, VV 124 may be the last bright Irr/Sph galaxy discovered in the
Local Group. This conclusion is based on the largescale studies of candidates for nearby galaxies con-
ducted by \citep{whi02,whi07}. The results obtained indicate with 77\% completeness that no more
than one or two galaxies with sizes $> 1\arcmin$ and surface brightnesses $< 25^m/\sq\arcsec$ in the $R$
 band can still be discovered in the Local Group.

Given the large sizes of VV 124 ($6.\arcmin5 \times 4.\arcmin2$) and its high surface brightness, the conclusions of Whiting
et al. (2007) may be considered to be unrelated to the probability of discovering such galaxies as VV 124.
This galaxy was not discovered previously as a member of the Local Group due to chance coincidences
of measurement errors \citep{kop08} and it is unlikely that such a coincidence of errors will be
repeated. The discovery of such a nearby galaxy is possible only in the Milky Way zone, because strong
extinction makes the small galaxies located there almost invisible.

The area of the Milky Way at extinction $A_B >$ 5 (corresponding to $A_I > $2.4) accounts for about 10\%
of the entire celestial sphere. In lower-extinction regions, a nearby galaxy would have already been found.
The Local Group galaxies form a fairly flat spatial structure \citep{pas07} and the probability
 of discovering a new galaxy within this flat structure is considerably higher than that in other
regions of the celestial sphere. Since the planes of the Milky Way and the Local Group do not coincide,
this reduces the probability of a possible new nearby galaxy being located in the Milky Way plane. Thus,
the presented data suggest that VV 124 can actually be the last bright Irr/Sph galaxy discovered in the
Local Group.

Based on long-slit spectra, we measured the radial velocities of two supergiants in VV 124. The
mean velocity estimated from these stars is $-86 \pm 15$\,km $^{-1}$ and the separation between them in projection
 onto the galaxy is 110 pc. The velocities of individual supergiants can have a dispersion of about
10 km s$^{-1}$ due to the random stellar motion; in addition, some of the stars may turn out to be binary,
which can introduce an error in measuring the radial velocity. For these reasons, the galaxy's velocity
can be determined more reliably from the velocity of unresolvable stars and diffuse gas. The velocity
of unresolvable stars in the galactic body measured from the combined spectrum of regions {\it a + b} on slit 1
(230 pc) and {\it b} on slit 2 (250 pc) is $-70 \pm 15$\,km s$^{-1}$.

The intensity centroid of unresolvable stars in these regions (given the slit positions) lies virtually in
the nuclear region, about 50 pc south of the galactic nucleus. The radial velocity of the gas with diffuse
[O III] emission on slit 2 is $-71 \pm 10$\,km s$^{-1}$; the region is localized in the western part at a distance
of about 150 pc from the nucleus. The velocity of the [O III] emission closer to the nucleus on the same
slit is $-47 \pm 15$\,km s$^{-1}$. The radial velocity of the diffuse [O III] emission on slit 1 is  $-54 \pm 15$\,km s$^{-1}$ .
The region of this emission is located at a distance of 100 pc southwest of the galactic center. Since
the velocity dispersion of the H\,I cloud that interacts with the galaxy is significant \citep{oos10}, we
can understand the noticeable velocity dispersion of the [O III] gas, i.e., possibly the same H\,I gas but
photoionized by young stars.

Analysis of our velocity measurements reveals no pattern related to the galactic rotation. Averaging the
velocities of all the components we measured (individual stars, the unresolvable stellar background, and
the [O III] gas emission) yields a mean velocity of   $-$69\,km s$^{-1}$, in good agreement with the measured
radial velocity of unresolvable stars. Based on these results, we conclude that the radial velocity of VV 124
is $-70 \pm 15$\,km s$^{-1}$.

The location of the region of the brightest young stars in VV 124 does not coincide with the center
of the galactic body \citep{kop08}. However, the region of young stars coincides with the brightest
part of the H\,I cloud discovered by Oosterloo (2010). The symmetry axis of the brightest stars coincides
in direction with that of the H\,I cloud. This asymmetry of young stars with respect to the galactic
body, their symmetry with respect to the H\,I cloud, which probably falls into the galaxy, and the young
stellar population in the central part of the galaxy (the dominance of hydrogen lines in the spectrum
of unresolved stars) are indicative of recent star formation processes. The star formation may be caused
by the interaction of the galaxy with H I clouds and, for this reason, we observe marked differences in the
velocities of the photoionized [O III] gas exceeding appreciably the measurement errors in different parts
of the galactic center ($-$47, $-$54, and $-$71\,km s$^{-1}$). A detailed analysis of 21-cm radio H\,I observations
 \citep{oos10} will clarify the internal dynamics of VV~124.

\section*{ACKNOWLEDGMENTS}

We wish to thank T. Oosterloo, who kindly provided the results of his 21-cm radio H I observations for
 VV 124 before their publication. We are also grateful to O.V. Mar'eva for help in reducing
the spectroscopic data. This work was supported by the Russian Foundation for Basic Research (project
nos. 09-02-00163 and 07-02-00909).

\begin{figure}[t]
\centerline{\includegraphics[angle=0, width=17 cm] {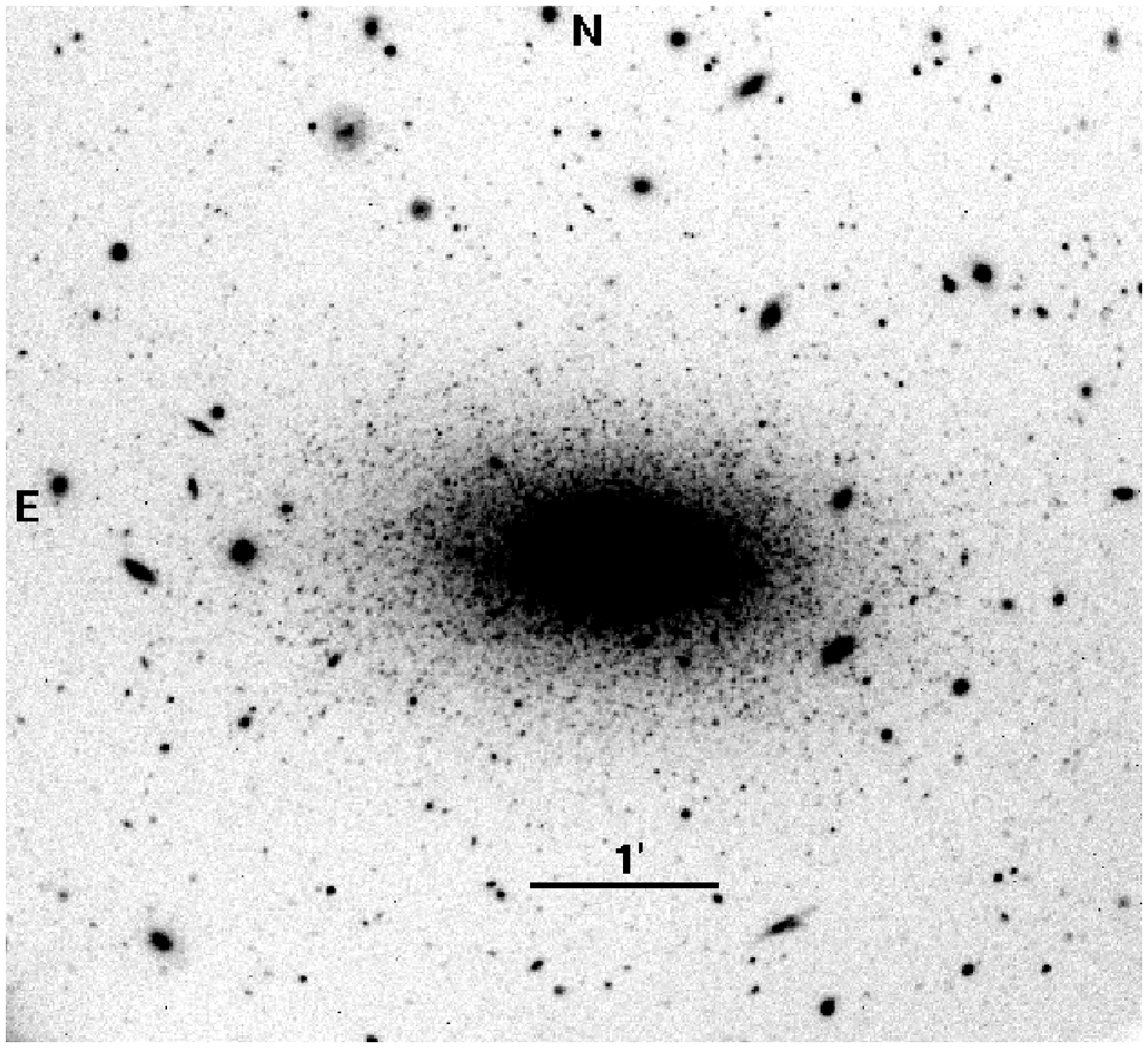}}
\caption{ $V$--band image of VV 124 (UGC 4879) obtained with the 6-m BTA telescope. We see that the galaxy is well resolved
into stars. Most of the faint stars in the image are red giants that form the thick galactic disk extending beyond the image field.}
\end{figure}

\begin{figure}[t]
\centerline{\includegraphics[angle=-90, width=10cm, bb=79 110 510 382,clip] {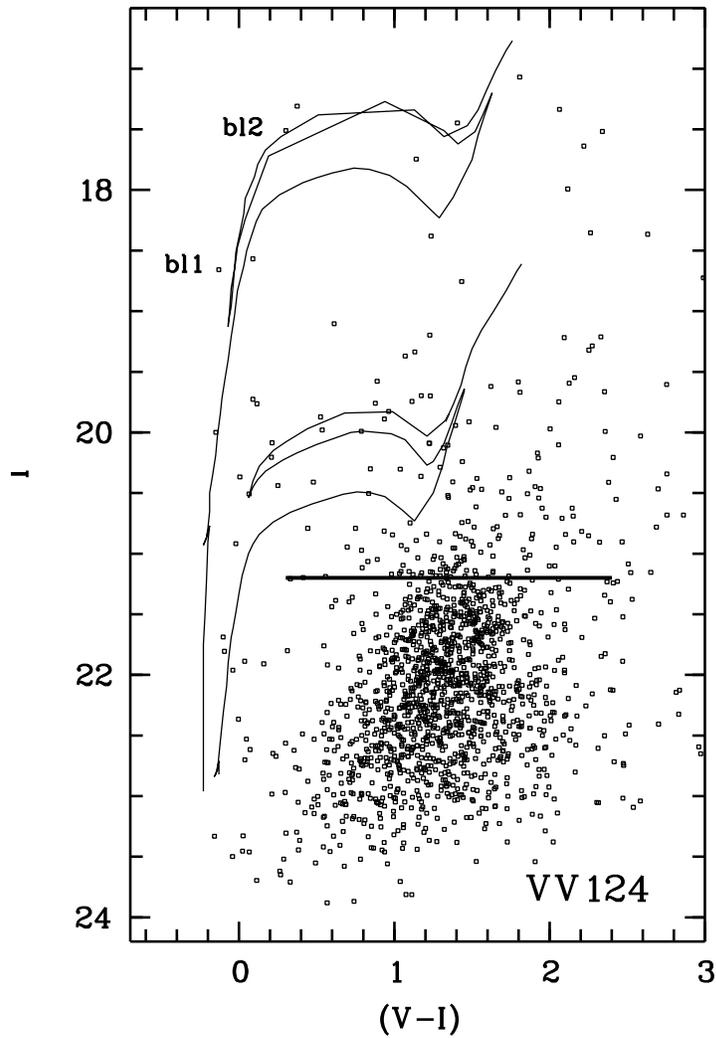}}
\caption {Hertzsprung--Russell (color--magnitude) diagram for stars in VV 124. We see a sparsely populated
blue supergiant branch and a large number of old stars, red giants that constitute the thick galactic disk. The
upper boundary of the red giant branch (TRGB jump, see also Fig. 3) is marked by the horizontal line. The
isochrones of stars with ages 16 (upper) and 63 (lower) Myr and metallicity Z = 0.008 are plotted on the diagram.
 We see from the isochrones that the marked bright supergiants bl1 and bl2 whose spectra are presented in
Fig. 6 are no older than 15--20 Myr.}
\end{figure}

\begin{figure}[t]
\centerline{\includegraphics[angle=-90, width=8cm, bb=186 133 453 382,clip] {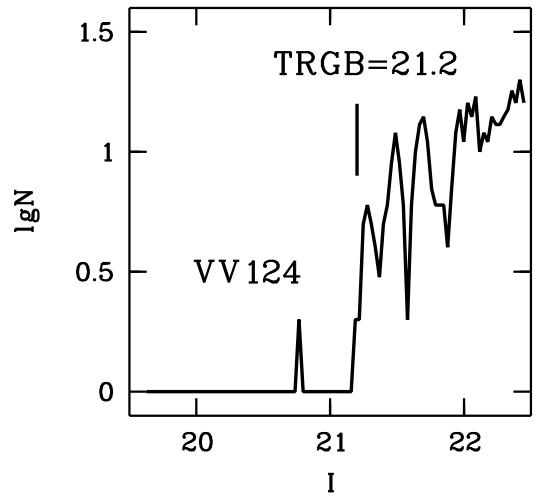}}
\caption{ Luminosity function for peripheral red giants in VV~124. At $I = 21.^m2$, the number of stars increases
sharply, which corresponds to the beginning of the RGB (TRGB jump).}
\end{figure}

\begin{figure}[t]
\centerline{\includegraphics[angle=-90, width=8cm, bb=186 65 533 388,clip] {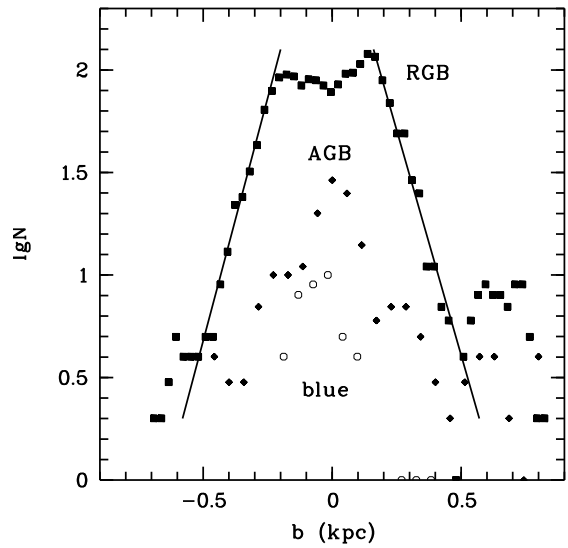}}
\caption{Number density distribution of stars with different ages (blue supergiants, AGB stars, red giants) along the
galaxy's minor axis. The number density of red giants decreases exponentially toward the galactic edge. A similar
distribution of stars is observed in the disks of irregular and low-mass spiral galaxies.}
\end{figure}

\begin{figure}[t]
\centerline{\includegraphics[angle=0, width=17cm] {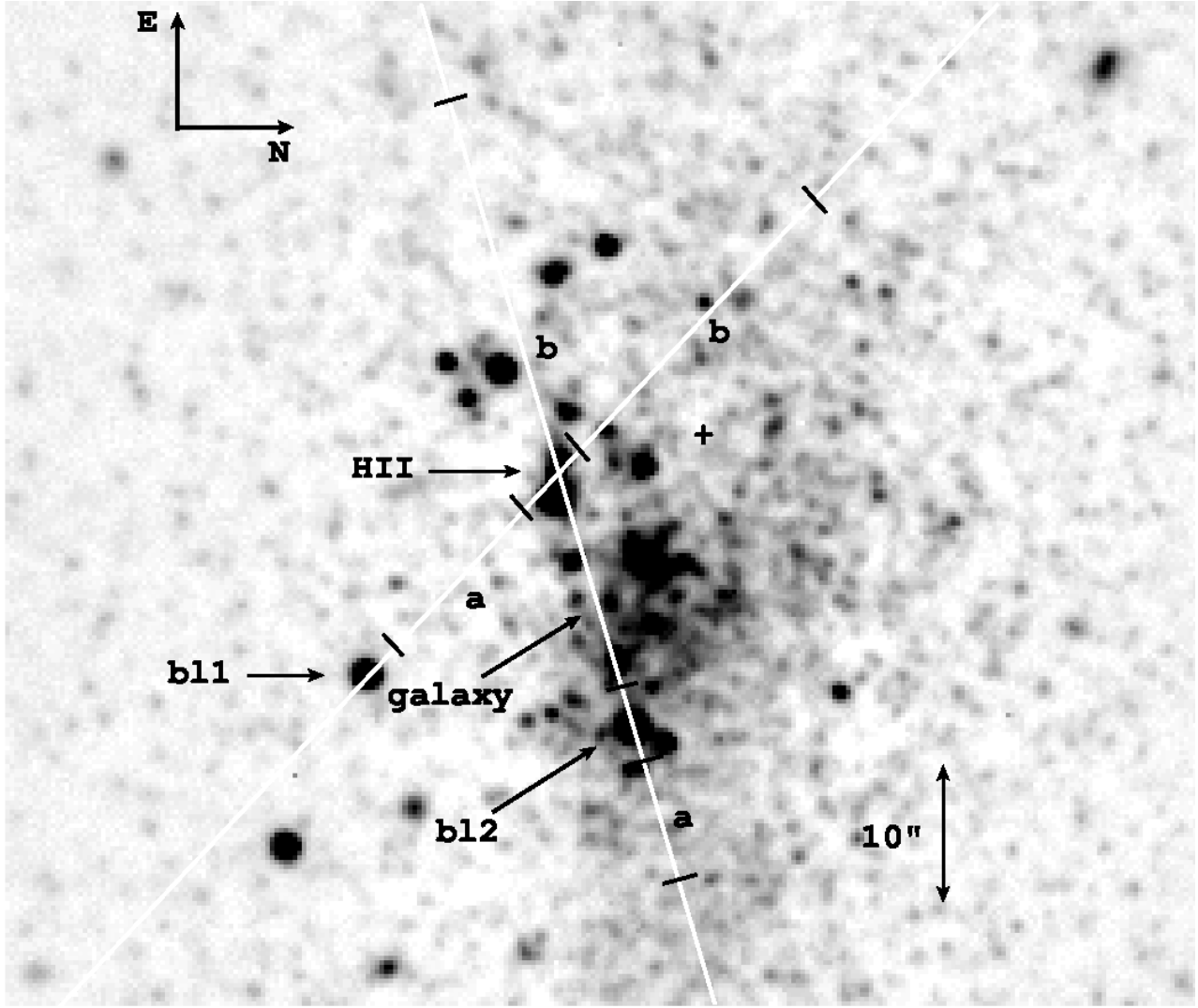}}
\caption{Image of the central part of VV~124 with the spectrograph slit positions marked. The supergiants bl1 and bl2, the
compact H II region, and the distant active galaxy that fell on slit 2 are marked. The spectrum of unresolved stars was extracted
in regions {\it a + b} of slit 1 and region b of slit 2. The diffuse [O III] emission was recorded in
 regions  {\it a} of slits 1 and 2 and in the western part of region {\it b} of slit 2.}
\end{figure}

\begin{figure}[t]
\centerline{\includegraphics[angle=0, width=10cm, bb=1 193 423 621,clip] {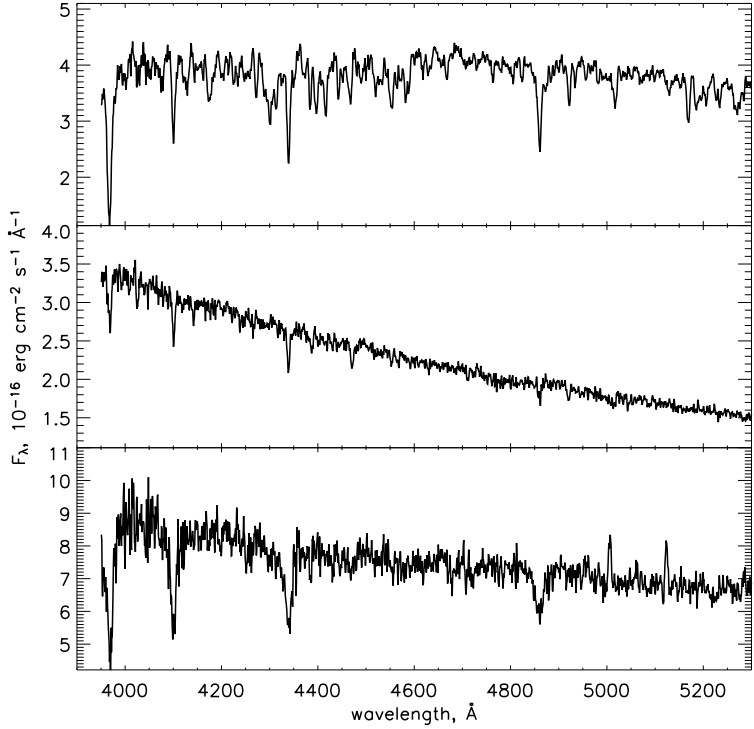}}
\caption {Spectra of the supergiant stars bl2 (a) with a temperature of $6000 \pm 100$\,K and bl1 (b) with a temperature of
$20000 \pm 3000$\,K and the spectrum extracted on slit bl2 (c) in the part of region {\it b}  13$\arcsec$ in length. 
The spectrum (c) was extracted from the western edge of region {\it b}  to the group of stars containing the compact 
H II region. In addition to the hydrogen absorption lines of the unresolved stellar background, 
the spectrum (c) exhibits the [O III] $\lambda$5007 emission line of the diffuse
high-excitation H II region and the emission line of the distant galaxy at $\lambda
\approx 5124$, presumably the [O II] \,$\lambda$  line. The 
Balmer H$\epsilon$, H$\delta$, H$\gamma$ and H$\beta$ lines are clearly seen in the spectrum of the stellar background.}
\end{figure}

\begin{figure}[t]
\centerline{\includegraphics[angle=0, width=17cm] {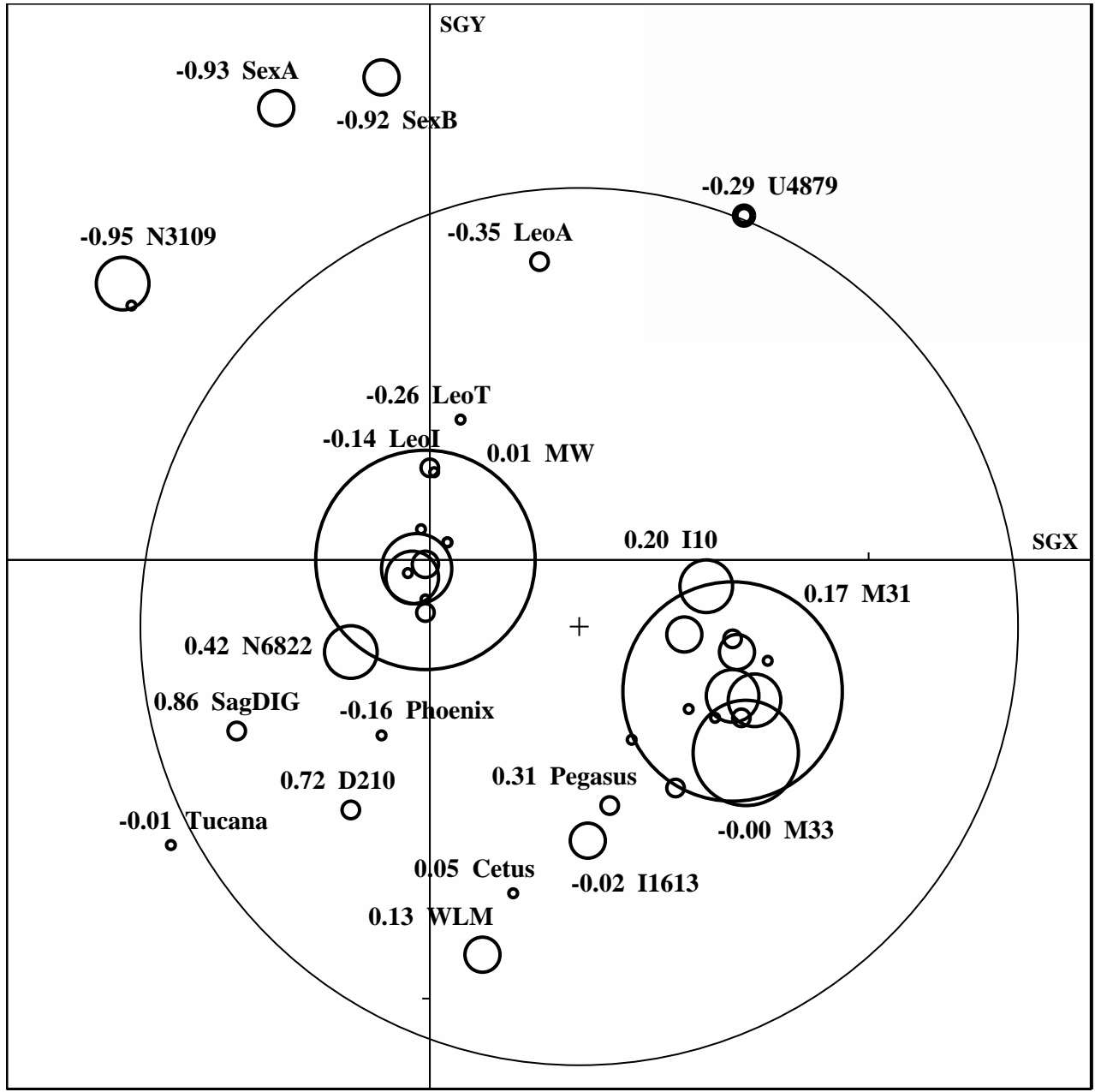}}
\caption{Distribution of Local Group and NGC~3109 group galaxies in projection onto the Supergalactic plane. The SGY axis is
directed upward and the SGX axis is directed rightward. The numbers indicate the SGZ coordinates. The large circle is 1 Mpc
in radius. The circles corresponding to our Galaxy (at the center) and M~31 are 0.25 Mpc in radius.}
\end{figure}

\begin{figure}[t]
\centerline{\includegraphics[angle=0, width=17cm] {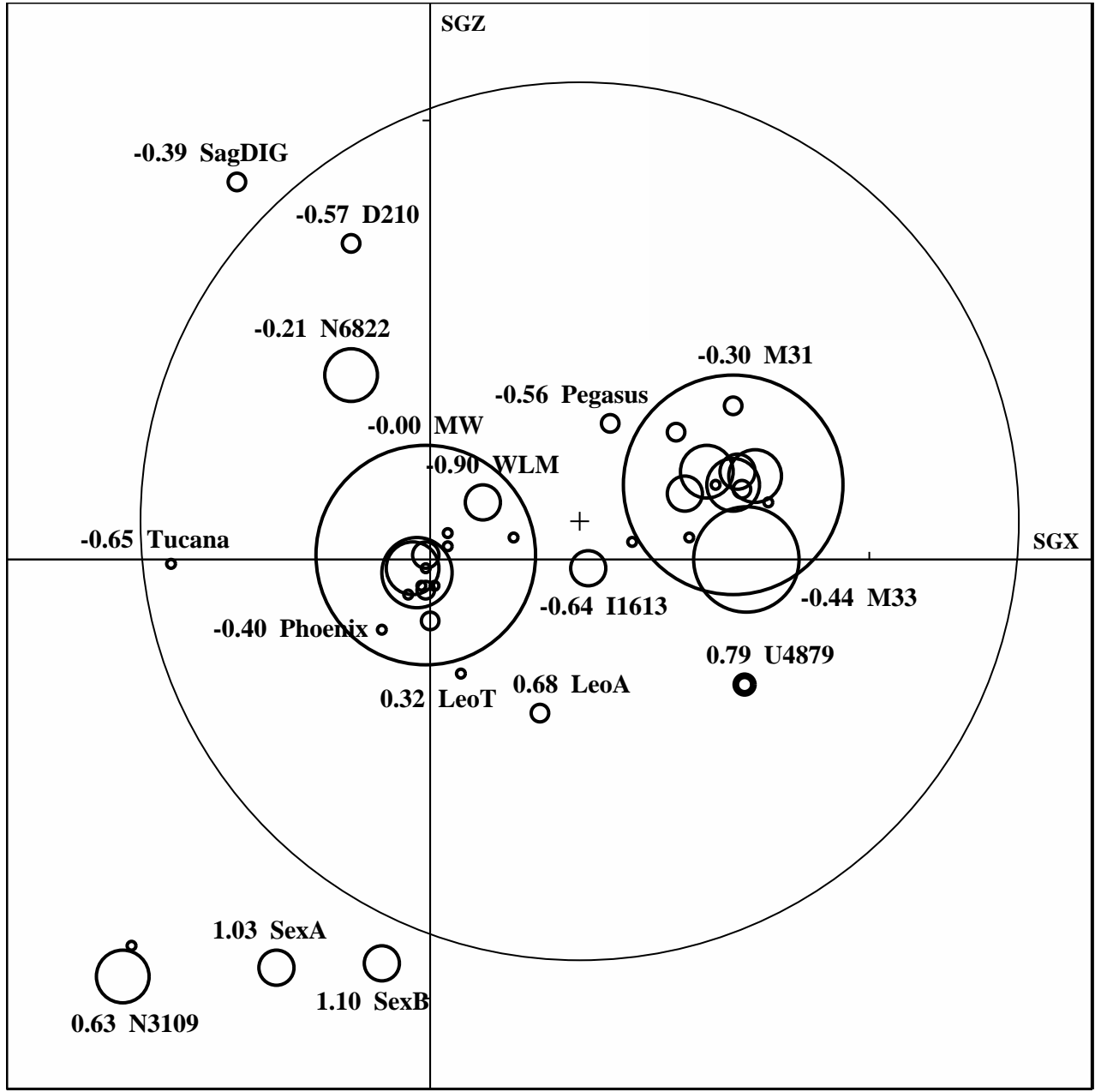}}
\caption{ Same as Fig. 7 in projection perpendicular to the Supergalactic plane. We see that VV~124 (UGC~4879) lies in the
plane of the Local Group disk together with Leo~A and Leo~T, while the nearby small group of galaxies (NGC~3109, Sex~A, and
Sex~B) greatly deviates from this plane.}
\end{figure}

\end{document}